\documentclass[runningheads]{llncs}
\usepackage{lineno}
\modulolinenumbers[5]
\usepackage[utf8]{inputenc}
\usepackage[T1]{fontenc}
\usepackage{cite}
\usepackage{amsmath,amssymb,amsfonts,mathrsfs}
\usepackage{algorithm,algorithmicx,algpseudocode,float}
\usepackage{graphicx}
\usepackage{array}
\usepackage[caption=false,font=footnotesize,labelfont=sf,textfont=sf]{subfig}
\usepackage{textcomp}
\usepackage[table,dvipsnames]{xcolor}
\usepackage{booktabs} 
\usepackage{multirow}
\usepackage{listings}
\def\BibTeX{{\rm B\kern-.05em{\sc i\kern-.025em b}\kern-.08em
    T\kern-.1667em\lower.7ex\hbox{E}\kern-.125emX}}

\newcommand{\blind}[1]{
\ifdefined\BLIND
{\color{gray}(omitted for blind review)}%
\else
{#1}
\fi
}

\usepackage{xspace}
\newcommand{\ie}{i.e.,\xspace}
\newcommand{\eg}{e.g.,\xspace}

\newcommand{\rv}{RISC-V\xspace}
\newcommand{\pabloplugin}{RAVE\xspace}
\xspaceaddexceptions{=}

\usepackage[acronym,toc,nonumberlist]{glossaries}
\makenoidxglossaries
\newacronym{fft}{FFT}{Fast Fourier Transform}
\newacronym{spmv}{SPMV}{Sparse matrix-vector multiplication}
\newacronym{hpc}{HPC}{High-Performance Computing}
\newacronym{csr}{CSR}{Control and Status Register}
\newacronym{isa}{ISA}{Instruction Set Architecture}
\newacronym{vl}{VL}{Vector Length}
\newacronym{pe}{PE}{Processing Elements}
\newacronym{epac}{EPAC}{European Processor Accelerator}
\newacronym{sdv}{SDV}{Software Development Vehicles}
\newacronym{fpga}{FPGA}{Field-programmable gate array}
\newacronym{ila}{ILA}{Integrated Logic Analyzer}
\newacronym{ovi}{OVI}{Open Vector Interface}
\newacronym{bfs}{BFS}{Breadth-First Search}
\newacronym{pr}{PR}{Page Rank}
\newacronym{cc}{CC}{Connected Compontents}
\newacronym{sssp}{SSSP}{Single Source Shortest Path}
\newacronym{td}{TD}{Top-Down}
\newacronym{bu}{BU}{Bottom-Up}
\newacronym{tf}{TF}{Threshold Factor}
\newacronym{tlb}{TLB}{Translation Lookaside Buffer}
\newacronym{rave}{RAVE}{RISC-V Analyzer for Vector Executions}
\newacronym{vpu}{VPU}{Vector Pocessing Unit}
\newacronym{bsc}{BSC}{Barcelona Supercomputing Center}
\newacronym{cuda}{CUDA}{Compute Unified Device Architecture}
\newacronym{sve}{SVE}{Scalable Vector Extension}
\newacronym{epi}{EPI}{European Processor Initiative}
\newacronym{simd}{SIMD}{Single-Instruction Multiple-Data}
\newacronym{hpcg}{HPCG}{High-Performance Conjugate Gradient}
\newacronym{gemm}{GEMM}{General Matrix-Matrix multiplication}

\usepackage{listings} 
\definecolor{col1}{HTML}{1E88E5}
\definecolor{col2}{HTML}{D81B60}
\definecolor{col3}{HTML}{43A047}
\definecolor{col4}{HTML}{F4511E}
\definecolor{col5}{HTML}{E259FF}

\usepackage{pifont}

\usepackage[colorlinks=true, allcolors=blue, bookmarks=false]{hyperref}

\RequirePackage[normalem]{ulem}
\RequirePackage{color}\definecolor{RED}{rgb}{1,0,0}\definecolor{BLUE}{rgb}{0,0,1}

\providecommand{\DIFdel}[1]{{\protect\color{RED}\sout{#1}}}
\providecommand{\DIFdel}[1]{}

\begin{document}

\title{
RAVE: RISC-V Analyzer of Vector Executions, a QEMU tracing plugin
}

\ifdefined\BLIND
\author{Authors omitted for blind review}
\else
\author{Pablo Vizcaino\inst{1}\orcidID{0000-0002-9253-8275} \and
Filippo Mantovani\inst{1}\orcidID{0000-0003-3559-4825} \and
Jesus Labarta\inst{1,2}\orcidID{0000-0002-7489-4727} \and
Roger Ferrer\inst{1}\orcidID{0000-0003-3306-8610}}
\fi
\institute{Barcelona Supercomputing Center, Barcelona 
\email{name.surname@bsc.es}
\and Universitat Politècnica de Catalunya
}

\maketitle

\begin{abstract}

Simulators are crucial during the development of a chip, like the \rv accelerator designed in the European Processor Initiative project.
In this paper, we showcase the limitations of the current simulation solutions in the project and propose using QEMU with \pabloplugin, a plugin we implement and describe in this document.
This methodology can rapidly simulate and analyze applications running on the v1.0 and v0.7.1 \rv V-extension.
Our plugin reports the vector and scalar instructions alongside useful information such as the vector-length being used, the single-element-width, and the register usage, among other vectorization metrics.
We provide an API used from the simulated Application to control the \pabloplugin plugin and the capability to generate vectorization traces that can be analyzed using Paraver.
Finally, we demonstrate the efficiency of our solution between different evaluated machines and against other simulation methods used in the \Gls{epac} project.

\end{abstract}


\section{Introduction: Motivation and state-of-the-art}\label{secIntro}

\Gls{hpc} is growingly adopting accelerated computing, with a common example being GP-GPU systems.
The Fugaku supercomputer~\cite{dongarra2020report}, which is the fourth most powerful supercomputer in the world according to the Top500 list (and first in the \Gls{hpcg} benchmark), takes another approach to acceleration through the usage of \Gls{simd} compute units.

The \Gls{simd} approach can be extended into vector computing, leveraging larger vectors with thousands of bits per register and more complex operations and memory access modes.
The NEC SX-Aurora vector Engine is a commercially available example of this acceleration approach, computing 256 double-precision elements per vector.
\rv is also leaning into vector computing through its recently ratified v1.0 V-extension.
One of the first implementators of the V-extension for \Gls{hpc} is the \Gls{epi} project, which includes the development of a \rv based accelerator with large vectors.

While the hardware is being produced, the \Gls{epi} project developed the \Gls{sdv}~\cite{mantovani_software_2023}, a set of software and hardware tools that facilitate the adoption of this technology before the complete hardware is available.
These tools include commercially available \rv platforms (without vector support) to develop the software ecosystem, compilers, \Gls{fpga}s to implement the accelerator's RTL, and the focus of this paper: simulators.
%
Software simulators are a crucial tool used in the development of new architectures, as they allow architects and software developers to perform a functional study of the architecture and they provide a testing field for the compiler.

%
Relevant work on simulating \rv processors has been researched in recent years, for example, using Static Binary Translation~\cite{lupori2018towards}.
Parallely, gemm5~\cite{binkert2011gem5} has been used to simulate \rv~\cite{roelke2017risc5}~\cite{zaourar2021multilevel}.
However, gemm5 focuses more on the architectural details and providing cycle-accurate measurements, at the cost of simulation times higher than other simulators such as QEMU~\cite{bellard2005qemu}.

We are interested in using QEMU, as our intention is not to perform cycle-accurate hardware simulations, but to analyze the vectorization of applications and kernels. 
%
%
QEMU has already been used to simulate \rv systems, for example to study \Gls{tlb} configurations~\cite{guo2019fast}, and while QEMU plugins have been used to count and analyze the simulated instructions~\cite{velea2017instruction}, to our best knowledge no such work has been done for the \rv V-extension.

When the \Gls{epi} project started in 2019, QEMU did not support the \rv V-extension, but we had scalar commercial \rv boards available.
The Vehave simulator was developed at \Gls{bsc} to take advantage of these boards.
It works by running \rv vector binaries on scalar cores natively until a vector instruction causes an Illegal Instruction signal (SIGILL).
Vehave captures this signal, simulates the vector instruction via software, and resumes the binary's execution on its next instruction.
s

Vehave lets us test the compiler's vectorization capabilities and verify the generated code's validity, and it generates execution traces with a granularity of vector instructions.
This allows us to profile how many instructions of each type are executed, with which vector-length, which datatypes are employed, which registers are used, and so forth.

Nevertheless, Vehave has three main disadvantages as a simulator.
First, it only has visibility on the vector instructions of the program, as it only runs after a SIGILL. 
The only way it can gather information on the scalar part of the code is by reading the hardware counters of the system.
However, even with this approach, significant noise gets mixed in the measurements due to the operating system's overhead on the SIGILL method and the Vehave overhead itself.

Secondly, for highly vectorized codes with large sequences of vector instructions, the program suffers the overhead of capturing the SIGILL, simulating, and going back to the user code, just to repeat this process on the next instruction, spending most of the runtime going back and forth through the operating system.

Thirdly, it lacks portability, as it requires that users have access to a \rv machine.
While the \Gls{sdv} systems inside the \Gls{epi} project include these machines, external third parties might prefer being able to run experiments on their own conventional systems or laptops.


Five years after the start of the \Gls{epi} project, QEMU supports the simulation of the \rv V-extension thanks to community efforts.
This opens the gates to replace the simulation software of the \Gls{sdv}s with QEMU, aiming to tackle the aforementioned issues of Vehave while keeping its strengths in the analysis methodology.
%


Additionally, QEMU allows users to write their own plugins, which hook into the program's simulation.
They do not alter the simulation but can monitor it down to the instruction granularity.
Some example uses of plugins include counting the number of executed instructions, memory accesses, or zones in the code that are repeatedly executed.

In this paper, we showcase how we modify QEMU to simulate the \gls{epac} chip, and we design a QEMU plugin called the \Gls{rave} to trace the application and collect vector metrics.

The rest of this paper is structured as follows:

Section~\ref{secImpl} explains the modifications to QEMU and the functionality of the \pabloplugin plugin. 
Section~\ref{secMet} introduces the hardware platforms where we compare the different simulation approaches. 
Section~\ref{secEval} presents the performance evaluation of our simulation technique with synthetic benchmarks and real applications.
Finally, we conclude the paper with our findings and comments in Section~\ref{secConc}.

\section{Plugin implementation}\label{secImpl}

In this section, we describe the small modifications we apply to QEMU, we define the metrics we want our plugin to collect during the simulation, we explain how the application can be instrumented to trace it, and then we describe the internals and flow of the \pabloplugin plugin.
The source files of the plugin, alongside scripts that set up a QEMU environment ready for \rv simulation, will be made public upon acceptance. 

\subsection{QEMU version and modifications}\label{secPatch}


When the \Gls{epi} project started, the currently ratified v1.0 version of the \rv V-extension did not exist yet, so the hardware was designed using the v0.7.1 version.
This means that we want our simulator to support both extensions, as we want to conduct tests that mimic the currently available hardware of the project but also start experimenting with the newer version. 
For this reason, we employ two different versions of the QEMU repository~\footnote{\url{https://gitlab.com/qemu-project/qemu/}}: commit $54e1f5be$ for v0.7.1 and commit $32b8913f$ for v1.0.


%
We perform two changes to the QEMU source code.
First, we edit the file \texttt{./target/riscv/cpu.h} so the \texttt{RV\_VLEN\_MAX parameter}, which defines the maximum bits per vector register, matches the \Gls{epac} value of 16384 bits.

\begin{figure}[!htbp]
\includegraphics[width=\textwidth]{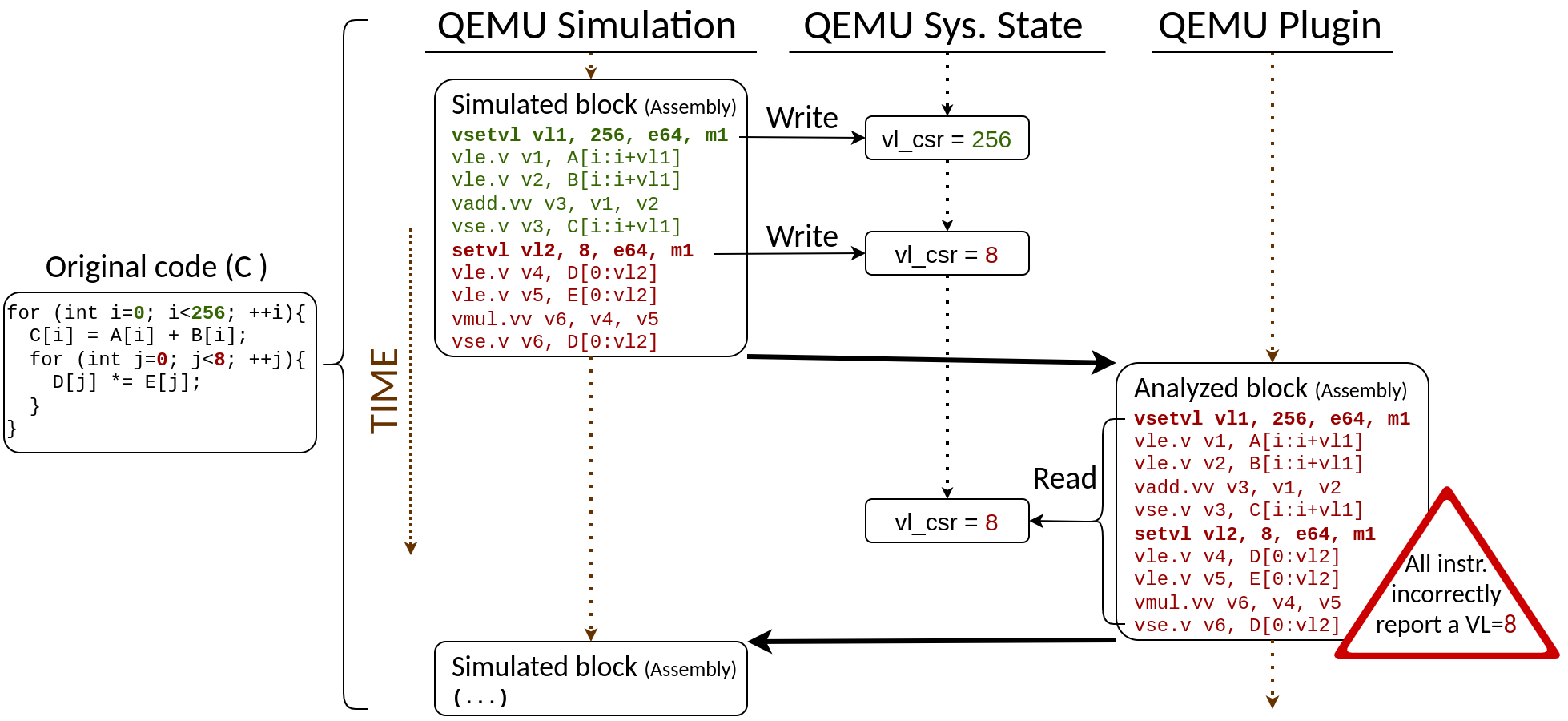}
\caption{Example diagram of inconsistent plugin reporting when the vector-length changes inside a simulated instruction block.}
\label{figInsn}
\end{figure}

The second change we apply requires a more detailed explanation of the functioning of QEMU.
Typically, QEMU works by simulating a block of instructions and then calling the plugin's hooks on all of them. 
This is good for performance, but does not allow to read a consistent state of the machine.
We show an example in Figure~\ref{figInsn}, where this mechanism does not consistently measure the vector-length of the simulated instructions.

We circumvent this issue by editing the file \texttt{./target/riscv/translate.c} so the \texttt{translator\_loop} function gets called with a \texttt{max\_insns} value of 1.
This conservative measure could be relaxed for zones where we are not required to read a consistent system state (\eg zones without vector instructions).

\subsection{Vectorization metrics and counters}\label{secMetrics}

The plugin we propose has to match these functionalities of Vehave:
\begin{itemize}
\item{Print the sequence of the executed vector instructions.}
\item{For each instruction, report the vector length, the element width, the LMUL, and the used registers.}
\item{Provide an API called for the application's code to define tracing regions.}
\item{Create an execution trace that can be opened using a \Gls{bsc} tool called Paraver~\cite{pillet1995paraver}.}
\end{itemize}

On top of these, we can add extra features not present in Vehave:
\begin{itemize}
\item{Count the exact number of scalar instructions between vector ones.}
\item{Define and count vectorization metrics on each instrumented block, such as the number of vector instructions depending on their type (arithmetic, memory, ...).}
\item{Generate a report at the end of the execution that summarizes these vectorization metrics.}
\end{itemize}

\begin{figure}[htpb!]
\includegraphics[width=\textwidth]{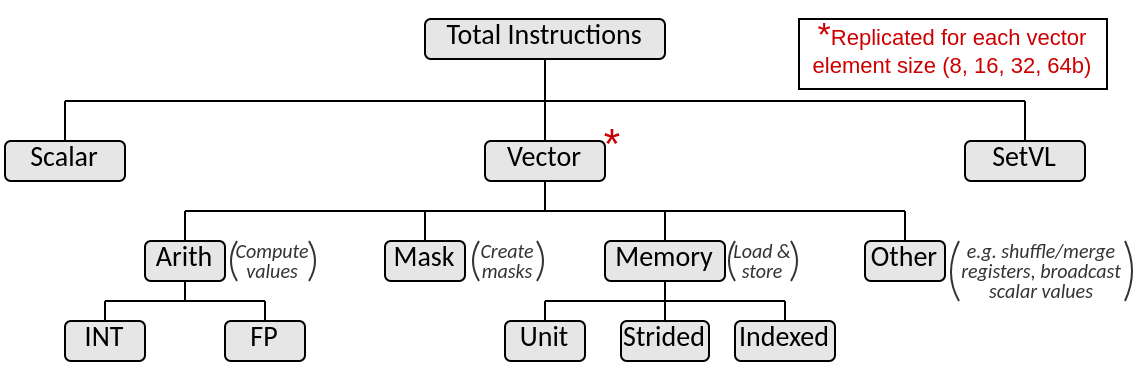}
\caption{Instruction classification used in the vectorization traces and reports generated by our \pabloplugin QEMU plugin}
\label{figClass}
\end{figure}

We present the classification we apply to the instructions in Figure~\ref{figClass}.

We can derive some metrics from these classes.
For example, dividing the Vector Instructions by the Total Instructions gives us the \textit{Vector Instruction Mix}.
We also keep track of the number of vector elements operated with all vector instructions, which can be divided by the number of Vector Instructions to get our code's \textit{Average Vector Length}.
We keep track of these counters defining the \texttt{qemu\_counters} structure shown in Figure~\ref{figCounters}.
\begin{figure}[htpb!]
\begin{lstlisting}[language=C]
#define SEWS 4
struct qemu_counters{
  double scalar_instr;
  double vsetvl_instr;
  double vector_instr[SEWS];
  double vunit_instr[SEWS];
  double vstride_instr[SEWS];
  double vidx_instr[SEWS];
  double vfp_instr[SEWS];
  double vint_instr[SEWS];
  double vmask_instr[SEWS];
  double velem[SEWS];
};
\end{lstlisting}
\caption{Structure holding the vectorization metrics counter by the \pabloplugin plugin.}
\label{figCounters}
\end{figure}

\subsection{Tracing mechanism}\label{secTrace}


Before diving into the plugin's description and execution flow, this section details how the simulated application can communicate with the plugin, instrumentating the code to generate traces.

The application cannot call plugin's functions, as QEMU is transparent to the application (\ie from its point of view, it is running natively). 
This means that our communication method needs to be encoded in the simulated instructions themselves.
However, these instructions should only appear when the user wants them; the compiler shall not generate them naturally.
For this reason, we use instructions that write to the $x0$ register, which is always hardcoded to the value 0, thus not used by the compiler under normal circumstances.

Specifically, we use instructions that load an immediate to a register (\texttt{li} and \texttt{lui}) so we can encode useful information in the immediate field.
Table~\ref{tabTrace} shows the three basic tracing functions that allow the user to enable and disable tracing, reducing the tracing overhead and the trace size.
\begin{table}[!htbp]
\caption{Simple tracing control using assembly instructions}
\begin{tabular}{|l|l|l|}
\hline
\multicolumn{1}{|c|}{\textbf{User Function}}    & \multicolumn{1}{c|}{\textbf{Instruction}} & \multicolumn{1}{c|}{\textbf{Description}} \\ \hline
qemu\_start\_trace()   & li x0, -3 & After this instruction, tracing is enabled   \\ \hline
qemu\_stop\_trace()    & li x0, -4 & After this instruction, tracing is disabled  \\ \hline
qemu\_restart\_trace() & li x0, -2 & Deletes tracing information up to this point \\ \hline
\end{tabular}
\label{tabTrace}
\end{table}

We also provide more advanced tracing mechanisms that let users define and name zones of the code, in a similar fashion to the Extrae \Gls{bsc} tool~\footnote{\url{https://tools.bsc.es/extrae}}, which works with tuples of events and values.
For example, an event might be the "Code Region", with values indicating different regions.
These tuples have numerical values, but they can be assigned names at the start of the execution, as shown in the code example on Figure~\ref{figExample}

\begin{figure}[htpb!]
\begin{lstlisting}[language=C]
int main(){
  double array1[256], array2[256], array3[256];
  qemu_name_event(1000,"Code Region")
  qemu_name_value(1000, 1, "Ini")
  qemu_name_value(1000, 2, "Compute")
  qemu_event_and_value(1000,1)
  ini_vectors(array1,array2,array3);
  qemu_event_and_value(1000,2)
  for(int i=0; i<256; ++i) array3[i] += array1[i] + array2[i];
  qemu_event_and_value(1000,0)
}
\end{lstlisting}
\caption{Example of the tracing API that the \pabloplugin plugin provides.}
\label{figExample}
\end{figure}

Table~\ref{tabEvent} shows how we encode events and values using assembly instructions.
Note that since the \texttt{qemu\_event\_and\_value} parameters may come from variables in the code not known at compile-time, so we cannot encode them in immediates.
We instead take advantage of the change proposed in Section~\ref{secPatch} that granted us a consistent state to read the contents of the simulated CPU registers.
%

\begin{table}[htbp]
\caption{Event and Value tracing mechanism implemented using assembly instructions}
\begin{tabular}{|l|l|p{0.6\linewidth}|}
\hline
\multicolumn{1}{|c|}{\textbf{User Function}}    & \multicolumn{1}{c|}{\textbf{Instruction}} & \multicolumn{1}{c|}{\textbf{Description}} \\ \hline
qemu\_event\_and\_value(e,v)                           & or x0, src1, src2   & event "e" and value "v" are read from src1 and src2. \\ \hline
\multirow{2}{*}{qemu\_name\_event(e,name)} & lui x0, e                            & The immediate specifies the event that gets named  \\ \cline{2-3} 
 &
  \begin{tabular}[t]{@{}l@{}}li x0, -1\\ lui x0, name{[}0{]},\\ lui x0, name{[}1{]}, ...\\ li x0, -1\end{tabular} &
  The "li" instructions mark the beginning and end of the name, and then a series of "lui" instructions encode its characters one by one \\ \hline
\multirow{2}{*}{qemu\_name\_value(e,v,name)} &
  \begin{tabular}[c]{@{}l@{}}lui x0, e\\ lui x0, v\end{tabular} &
  The imm. "v" specifies the value that gets named, and "e" the event that contains that value \\ \cline{2-3} 
 &
  \begin{tabular}[c]{@{}l@{}}li x0, -1\\ lui x0, name{[}0{]},\\ lui x0, name{[}1{]}, ...\\ li x0, -1\end{tabular} &
  The value name is transmitted using the same protocol as the event name. \\ \hline
\end{tabular}
\label{tabEvent}
\end{table}

\subsection{\pabloplugin description}\label{secPlugin}


The plugin starts by hooking a function to the translations of instruction blocks (which we defined to be of size $1$ in Section~\ref{secPatch}).
This function performs three main tasks, as shown in the pseudocode in Algorithm~\ref{algTrans}.

\begin{algorithm} [htbp]
\caption{Pseudocode of the plugin's translation block function callback}
\label{algTrans}
\begin{algorithmic}[1]
\Procedure{vcpu\_tb\_trans}{translation\_block}
\For{$instr \in translation\_block$}
\State{String disas $\gets$ qemu\_plugin\_insn\_disas(instr)}
\If {type(disas) == scalar}
\State{set\_callback(vcpu\_insn\_exec, generic\_scalar\_data)}
\ElsIf {type(disas) == vector}
\State{instr\_data $\gets$ fill\_vec\_struct(instr)}
\State{set\_callback(vcpu\_insn\_exec, instr\_data)}
\ElsIf {type(disas) == tracing}
\State{func $\gets$ specific\_tracing\_function(instr)}
\State{set\_callback(func, instr\_data)}
\EndIf
\EndFor
\EndProcedure
\end{algorithmic}
\end{algorithm}

%
%
First, the function loops through the instructions in the translation block, and disassembles them into a string.
QEMU can disassemble instructions, but this feature is only available for the v1.0 V-extension.
For v0.7.1, we implement our own disassembler as a $C$ function following the official specification~\footnote{\url{https://github.com/riscv/riscv-v-spec/releases/tag/0.7.1}}.


Once the instruction is disassembled, we perform different tasks depending on the instruction's type: vector, scalar, and tracing.
Tracing instructions, previously explained in Section~\ref{secTrace}, are specific instructions added in the simulated application to communicate with the plugin.
Vector instructions instantiate the structure shown in Figure~\ref{figStruct} with the information we want to trace, including the instruction classification method previously depicted in Figure~\ref{figClass}.

\begin{figure}[htbp!]
\begin{lstlisting}[language=C]
enum instr_type{SCALAR, VECTOR, VSETVL};
enum v_major_type{OTHER, ARITH, MEMORY, MASK};
enum v_minor_type{NOTYPE, FP, INT, UNIT, STRIDE, INDEX};
struct instr_data{
  uint64_t PC; 
  uint32_t paraver_code; //Needed by BSC trace visualization system
  char * asm_string; // E.g. vadd.vv v0, v3, v4
  short dst, src1, src2, src3
  enum instr_type type;
  enum v_major_type v_majortype;
  enum v_major_type v_minortype;
};
\end{lstlisting}
\caption{Instruction data kept for each simulated vector instruction.}
\label{figStruct}
\end{figure}

Scalar instructions do not keep track of this information, reducing the simulation time and trace size, since in most use cases counting the number of scalar instructions between vector instructions gives us the desired insight (\ie we do not need to know their assembly code or used registers). 
However, the user can also tell QEMU to record this scalar information at the cost of larger simulations.


Finally, a callback function is assigned to each instruction, which gets called when they are executed. 
Vector and scalar instructions use the same callback (with the instruction structure as parameter), incrementing the counters defined in Section~\ref{secMetrics} accordingly.
The callback also logs and traces vector instructions.

The tracing instructions get their own callback function depending on their function, described before in Table~\ref{tabTrace} and Table~\ref{tabEvent}.
%
%
Specifically, the \texttt{or} instruction generated by \texttt{qemu\_event\_and\_value(x,y)} has a callback that reads the scalar registers containing the event and value identifiers.
It also updates a list of \textit{execution regions}, which are defined as the instructions enclosed between two calls to \texttt{qemu\_event\_and\_value(x,y)} with the same event.
When this callback opens a new \textit{execution region}, the vectorization counters are read and stored in the region structure. 
When the region is closed, its counters are computed substracting the current ones minus the region starting counters.
Once the simulation finishes, the plugin prints the contents of the \textit{execution region}.

\begin{figure}[!htbp]
\includegraphics[width=\textwidth]{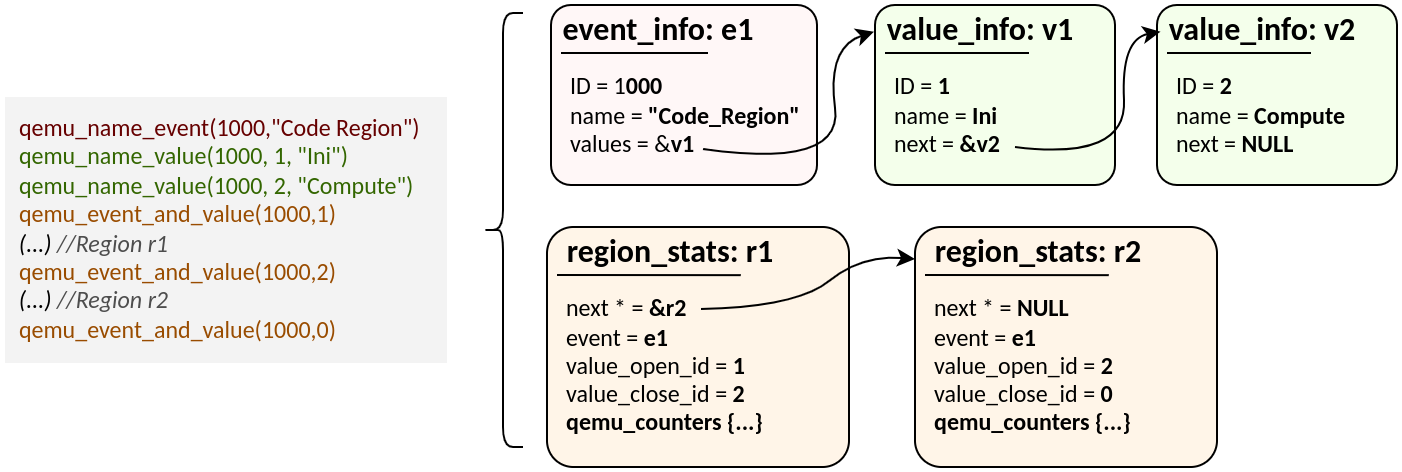}
\caption{Example of data structures created to track events, values, and regions.}
\label{figStructures}
\end{figure}

Figure ~\ref{figStructures} depichts this process and the auxiliary data structures needed to keep track of events and values.
In the example, the first \texttt{qemu\_event\_and\_value} opens region \textit{r1}, the second one closes \textit{r1} and opens \textit{r2}, and the third one closes \textit{r2}.
This way, \textit{r1} saves the vectorization counters encapsulated between the values $1$ and $2$ of event $1000$, and \textit{r2} saves the countes between $2$ and $0$.
 
\section{Evaluated Platforms}\label{secMet}

After implementing the \pabloplugin plugin, we evaluate it and benchmark its performance on four machines:
%
%

\textbf{Unmatched:}
First, we compare the performance achieved with QEMU against the previous \Gls{sdv} simulator, Vehave, running on HiFive Unmatched~\footnote{\url{https://www.sifive.com/boards/hifive-unmatched}} scalar \rv boards. 
These include a quad-core, 64-bit processor at $1.2GHz$, with 16GB of DDR4 memory, running an Ubuntu 22.04.2 LTS operating system.

\textbf{AMD:}
We run QEMU on one of the x86 machines available in our laboratory, which has the $12$-core AMD Ryzen 5 5600G processor running at $3.9Ghz$, with 32GB of DDR4 memory, on top of an Ubuntu 20.04.4 LTS operating system.

\textbf{Laptop:}
We also use a medium-end laptop to prove the portability of our solution.
This laptop has an Intel(R) Core(TM) i7-8650U CPU 8-core CPU running at $2.1GHz$, with 16GB of DDR4 memory and an Ubuntu 22.04.4 LTS.

\textbf{EPAC@FPGA:}
Finally, we compare our software-simulation results against the hardware emulation of the \Gls{epac} chip inside an \Gls{fpga} running at $50MHz$.
The \Gls{fpga} board is the Virtex UltraScale+ HBM VCU128 FPGA Evaluation Kit\footnote{\url{https://www.xilinx.com/products/boards-and-kits/vcu128.html}}. 
This board includes a VU37P FPGA\footnote{Complete device name: XCVU37P-2FSVH2892E} with 8~GB of integrated HBM memory and 4.5~GB of DDR4 memory. 

\section{\pabloplugin evaluation}\label{secEval}

In this section, we analyze the performance of our plugin and show its potential for vectorization analysis.
We perform two types of benchmarks: synthetic and real applications. 
All codes are compiled using an LLVM-based compiler\footnote{\blind{\url{https://repo.hca.bsc.es/gitlab/rferrer/llvm-epi}}} developed at \Gls{bsc} capable of vectorizing code for the v1.0 and v0.7.1 V-extension of \rv.
We want to highlight that the same compiled binary runs seamlessly on the four platforms described in Section~\ref{secMet} undergoing three different types of execution (QEMU simulation, Vehave simulation, and FPGA emulation).

\subsection{Benchmarking}\label{secPerf}

\begin{figure}[!htbp]
\includegraphics[width=\textwidth]{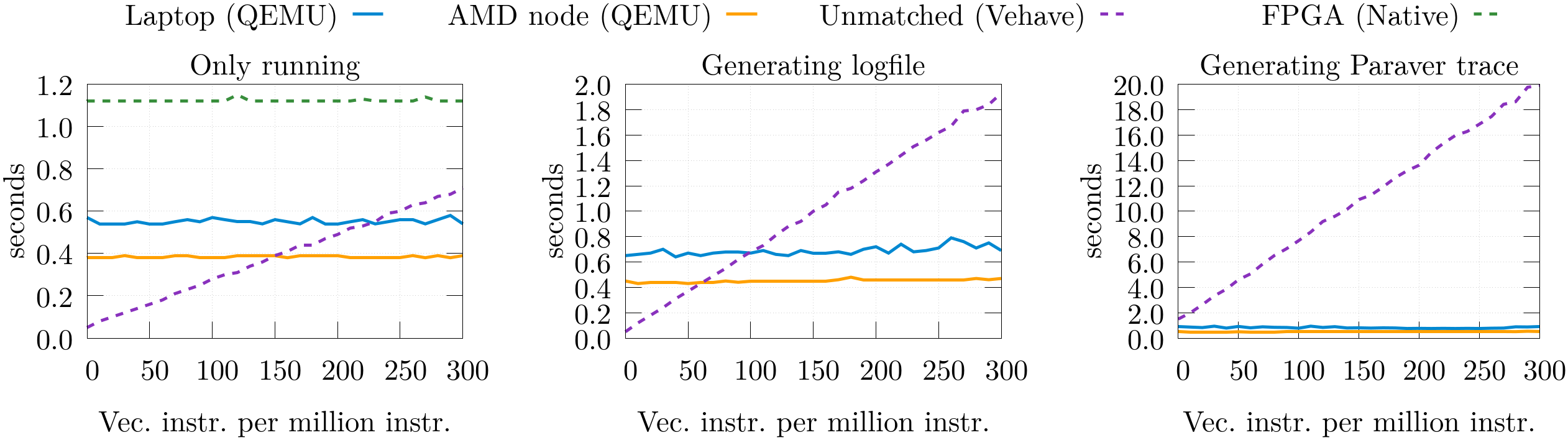}
\caption{Simulation time depending on the number of vector instructions per million total instructions, depending on the simulation method and platform.}
\label{figBench}
\end{figure}

First, we run a synthetic benchmark that executes $i_v$ vector instructions, $i_s$ scalar instructions, and $i_t$ total instructions ($i_{t} = i_{v} + i_{s}$).
We adjust the ratio $r_{v} = \frac{i_v}{i_t}$ to study how higher vectorization affects the simulation time.
Figure~\ref{figBench} shows the result of this benchmark with $i_{t} = 10^{8}$, the horizontal axis showing $r_{v}$ in vector instructions per million total instructions.
Dashed lines are used to represent non-QEMU simulations.

We show three distinct experiments: One only running the simulation, another generating a log file with the executed instructions, and another generating a Paraver trace.
The \Gls{fpga}-based simulation does not have visibility of the individual instructions, so it is only present in the first experiment.
\Gls{fpga} and QEMU-based simulation are not affected by the $r_v$ ratio in any experiment, with the AMD node generally being the fastest solution.
With an extremely low number of vector instructions (less than $0.02\%$), the Vehave simulator running on the Unmatched board is the fastest method unless we generate a Paraver trace, which takes significantly more time than QEMU.


The second type of test we perform is running scientific kernels vectorized at \Gls{bsc} with the different simulation methods and evaluated platforms. 
We run four graph algorithms from the libPVG vector library~\cite{vizcaino2024graphs}, specifically \Gls{bfs}, \Gls{pr}, \Gls{cc}, and \Gls{sssp}, using an input graph of 16 thousand nodes.
We also run a vectorized \Gls{fft}~\cite{fftp}, performing a one-dimensional forward transform of half a million complex elements, a vectorized \Gls{gemm} and a vectorized \Gls{spmv}~\cite{gomez2021efficiently}.

\begin{figure}[!htbp]
\includegraphics[width=\textwidth]{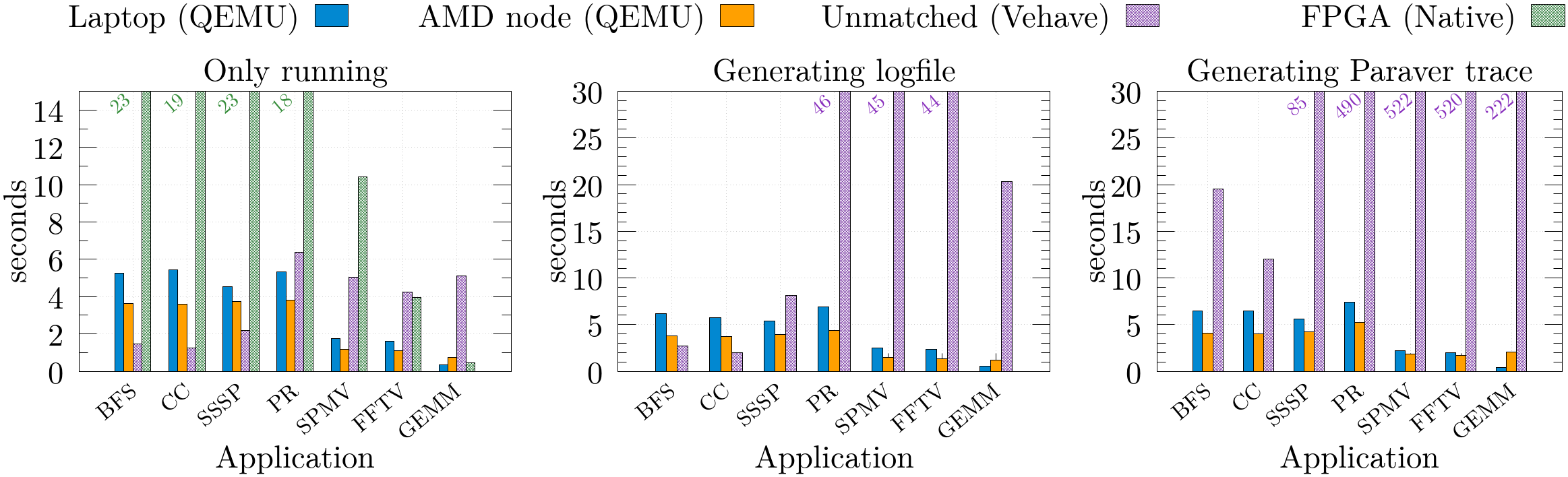}
\caption{Simulation time on different HPC kernels depending on the simulation method and platform.}
\label{figApps}
\end{figure}

Figure~\ref{figApps} shows the execution time of these applications under the different simulation methods and platforms.
We can see that for the four graph algorithms, Vehave is the fastest method, while for the other three codes, QEMU is significantly faster, especially for the \Gls{gemm}.

We can correlate these results with the synthetic benchmark on Figure~\ref{figBench}.
The graph algorithms spend most of the execution time reading the graph from disk (a completely scalar process), so it makes sense that Vehave can take advantage of the native execution of scalar instructions to be the fastest solution.
\Gls{spmv} and \Gls{fft} both have a scalar initialization phase, but it is not large enough to make Vehave faster than QEMU.
Finally, the instructions in \Gls{gemm} are mostly vector, thus showing the greatest time difference in favor of QEMU.

\subsection{Use case: Analyzing a Graph algorithm}\label{secUse}

\begin{figure}[htbp!]
\includegraphics[width=\textwidth]{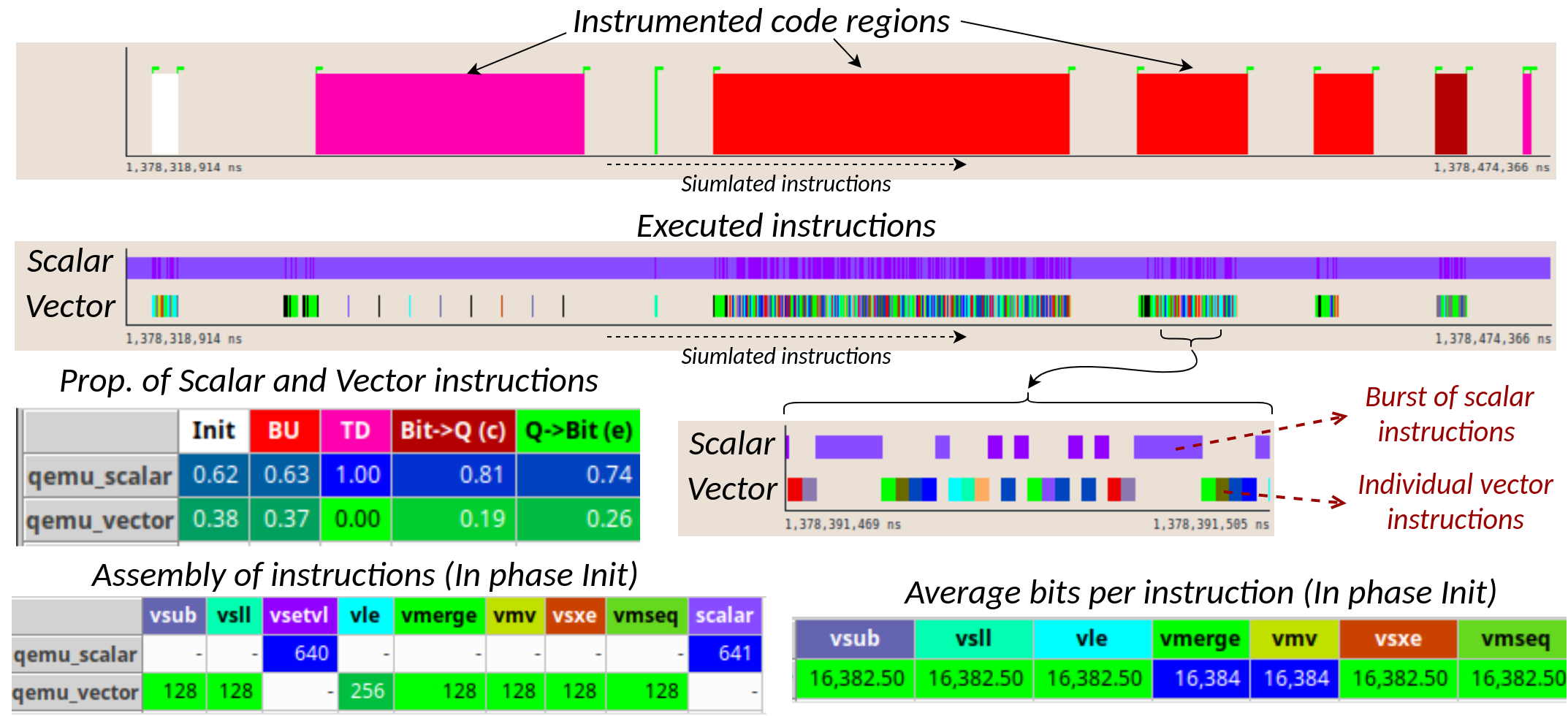}
\caption{Paraver visualization of a QEMU simulation of BFS.}
\label{figParaver}
\end{figure}

In this section, we exemplify the type of insight that the \pabloplugin plugin can give to application, compiler, and hardware developers.
In Figure~\ref{figParaver}, we use Paraver to visualize a simulation trace of \Gls{bfs}.
At the top we see two sequences where the horizontal axis represents the simulated instructions.
The first view shows the code's instrumented regions using different colors.
The view under it uses colors to differentiate vector and scalar instructions in two separate lines, which can be seen better in the zoom-in view.
Note that scalar instructions all use the same color, as we do not differentiate between them to reduce the trace size.
%

We also plot three tables obtained with Paraver, showing the proportion of scalar and vector instructions per each phase, the instructions count by their assembly code (for the first code phase), and the average bits they operate on.
We can see that most phases vectorize correctly (around 30\% of vector instructions), with high vector length, but the TD phase (in pink) is nearly completely scalar.

\begin{figure}[htbp!]
\includegraphics[width=\textwidth]{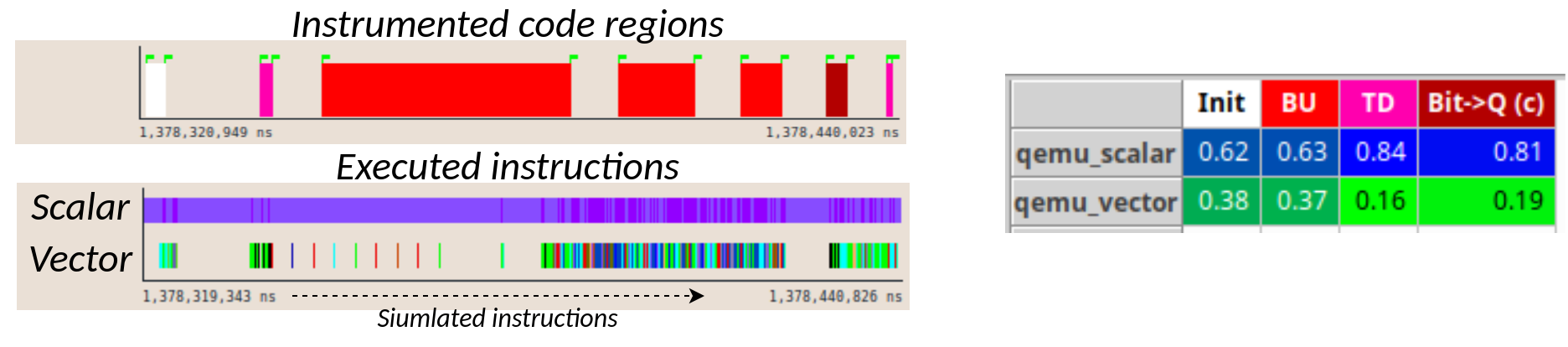}
\caption{Paraver visualization of a QEMU simulation of BFS after increasing TD vectorization.}
\label{figTD}
\end{figure}

After applying some changes to the code, we managed to increase the vectorization of the TD phase, and the results can be seen in Figure~\ref{figTD}.

We can also visualize this information on the terminal where we execute the QEMU simulation. 
At the left of Figure~\ref{figConsole}, you can see the plugin's output for the first BU region of the \Gls{bfs} simulation.
A significant percentage of the vector instructions are of type \textit{"Mask"} or \textit{"Other"}, which, in general, we should avoid or reduce, as they do not compute or load data. 
At the right of Figure~\ref{figConsole}, we show the same report for a newer code version that removes some control flow from the vectorized code, reducing the need for masks.

\begin{figure}[!htbp]
\begin{minipage}{.5\textwidth}
\center{Snippet of the report before BFS control-flow optimization}
\begin{lstlisting}
Reg. #3: Event 1000(code_region), Value 3(BU)
    tot_instr: 38872
        scalar_instr: 15818 (40.69 %)
        vsetvl_instr: 5236 (13.47 %)
        SEW 64 vector_instr: 17818 (45.84 %)
            avg_VL: 255.60 elements
            Arith: 2466 (13.84 %)
                FP: 0 (0.00 %)
                INT: 2466 (100.00 %)
            Mem: 3142 (17.63 %)
                unit: 1573 (50.06 %)
                strided: 0 (0.00 %)
                indexed: 1569 (49.94 %)
            Mask: 8171 (45.86 %)
            Other: 4039 (22.67 %)
\end{lstlisting}
\end{minipage}
\begin{minipage}{.5\textwidth}
\center{Snippet of the report after BFS control-flow optimization}
\begin{lstlisting}
Reg. #3: Event 1000(code_region), Value 3(BU)
    tot_instr: 44780
        scalar_instr: 21866 (48.83 %)
        vsetvl_instr: 9556 (21.34 %)
        SEW 64 vector_instr: 13358 (29.83 %)
            avg_VL: 254.77 elements
            Arith: 2481 (18.57 %)
                FP: 0 (0.00 %)
                INT: 2481 (100.00 %)
            Mem: 3028 (22.67 %)
                unit: 1454 (48.02 %)
                strided: 0 (0.00 %)
                indexed: 1574 (51.98 %)
            Mask: 4992 (37.37 %)
            Other: 2857 (21.39 %)
\end{lstlisting}
\end{minipage}
\caption{Console report of the QEMU plugin for the BU region of the \Gls{bfs} simulation, before and after a code optimization.}
\label{figConsole}
\end{figure}

As seen when comparing both reports in Figure~\ref{figConsole}, the number of "Mask" and "Other" instructions decreases.
In both versions, we can also observe that half of the vector memory instructions are unit-strided and the other half indexed.
This insight is useful for the hardware architects designing the \Gls{epac} chip, as it hints that dedicating efforts to implement indexed memory operations efficiently will substantially benefit the performance of this kind of applications.

\section{Conclusions}\label{secConc}

In this paper, we introduced the need for fast and insightful simulators during the development of a chip, focusing our scope on the \Gls{epac} design in the \gls{epi} project. 
We exposed some drawbacks of the current simulation tools present in the \gls{sdv} methodology, and proposed a solution using QEMU.

The \pabloplugin plugin implemented and described in this paper has proven to be capable of rapidly producing vectorization reports for the \rv V-extension v1.0 and v0.7.1.
Our plugin reports the vector and scalar instructions simulated, alongside information such as the vector length and other vectorization metrics.
We provide an API that can be called from within the simulated application, and the capability to generate traces that can be visualized using Paraver.

When evaluating the performance of our plugin between different machines and against other simulation methods, such as Vehave or \gls{fpga} hardware emulation, we showed that QEMU is the fastest solution when the simulated code has vector instructions.
Vehave, the old simulation method of the \Gls{sdv}, is only faster when the code is virtually completely scalar and we do not generate a report.
Finally, we also exemplify a use-case of the utility of the plugin's reports and traces using a vectorized graph computing application.

The next steps of this plugin include doing a more in-depth performance analysis to study where the simulation time is being spent, and analyze the memory footprint of the plugin.
Limiting the simulation block to one instruction allowed us to read a consistent state from the plugin, but future iterations might benefit from a flexible block size (\eg only limited to one instruction in vectorized zones).
We will also work tightly with the performance analysts who are starting to use this tool, gather feedback, and implement more tracing and reporting functionalities, such as more detailed metrics on memory instructions (\eg load/store differentiation, stride analysis, or indexed pattern recognitions).
Finally, the \Gls{epac} project will soon advance to multi-core designs, so validating the behavior of vector codes on multiple cores using QEMU will be paramount.

\section*{Acknowledgment}

\blind{Supported by the EuroHPC Joint Undertaking (JU): FPA N. 800928 (EPI) and SGA N. 101036168 (EPI-SGA2). The JU receives support from the EU Horizon 2020 research and innovation programme and from Croatia, France, Germany, Greece, Italy, Netherlands, Portugal, Spain, Sweden, Denmark and Switzerland. The EPI-SGA2 project, PCI2022-132935 is also co-funded by MCIN/AEI /10.13039/501100011033 and by the UE NextGenerationEU/PRTR.}

\bibliographystyle{splncs04}
\bibliography{99-bib}

\end{document}